\documentclass[10pt,onecolumn]{aastex631}
\hypersetup{linkcolor=magenta, citecolor=cyan, filecolor=yellow, urlcolor=blue}
\usepackage{indentfirst}
\usepackage{comment}
\usepackage{multirow} 
\usepackage{booktabs}
\usepackage{bigints}
\usepackage{mathrsfs,amsmath}

\newcounter{tr}
\ifnum \value{tr}>5
\newcommand{\deletedD}[1]{{\color{red} Damien - Deleted: } \sout{#1}}

\newcommand{\authorcommentD}[1]{{\color{red} Damien - Comment :} {\color{blue} #1}}

\else
\newcommand{\deletedD}[1]{}

\newcommand{\authorcommentD}[1]{}

\fi
\begin{document}

\title{Is magnetically dominated outflow required to explain GRBs?}

%\author{Damien B\'egu\'e$^1$, Liang Li$^2$ and Gregory Vereshchagin$^3$}
%\address{$^1$ Department of Physics, Bar Ilan University, Ramat-Gan 52900, Israel \\ Max-Planck-Institut f\"ur extraterrestrische Physik, Giessenbachstrasse, D-85748 Garching, Germany \\
%$^2$ ICRA and Dipartimento di Fisica, Universit\`a  di Roma ``La Sapienza'', Piazzale Aldo Moro 5, I-00185 Roma, Italy and \\
%International Center for Relativistic Astrophysics Network, Piazza della Repubblica 10, I-65122 Pescara, Italy and \\
%INAF, Viale del Parco Mellini 84, 00136 Rome, Italy \\
%$^3$ ICRA and Dipartimento di Fisica, Universit\`a  di Roma ``La %Sapienza'', Piazzale Aldo Moro 5, I-00185 Roma, Italy and \\
%International Center for Relativistic Astrophysics Network, Piazza della Repubblica 10, I-65122 Pescara, Italy and \\
%INAF, Istituto di Astrofisica e Planetologia Spaziali, Via Fosso del Cavaliere 100, 00133 Rome, Italy}
%\date{November 2021}

\author[0000-0002-1623-3576]{Gregory Vereshchagin}
\affiliation{ICRANet, Piazza della Repubblica 10, I-65122 Pescara, Italy}
\affiliation{ICRA and Dipartimento di Fisica, Universit\`a  di Roma ``La Sapienza'', Piazzale Aldo Moro 5, I-00185 Roma, Italy}
\affiliation{INAF, Istituto di Astrofisica e Planetologia Spaziali, Via Fosso del Cavaliere 100, 00133 Rome, Italy}

\author[0000-0002-1343-3089]{Liang Li}
\affiliation{ICRANet, Piazza della Repubblica 10, I-65122 Pescara, Italy}
\affiliation{ICRA and Dipartimento di Fisica, Universit\`a  di Roma ``La Sapienza'', Piazzale Aldo Moro 5, I-00185 Roma, Italy}
\affiliation{INAF -- Osservatorio Astronomico d'Abruzzo, Via M. Maggini snc, I-64100, Teramo, Italy}

\author[0000-0003-4477-1846]{Damien B\'egu\'e}
\affiliation{Department of Physics, Bar Ilan University, Ramat-Gan 52900, Israel}

%\correspondingauthor{Damien B\'egu\'e; Liang Li; Gregory Vereshchagin}
%\email{cayley38@gmail.com; liang.li@icranet.org; veresh@icra.it}
\correspondingauthor{Gregory Vereshchagin}
\email{veresh@icra.it; liang.li@icranet.org; begueda@biu.ac.il }

\begin{abstract}
The composition of relativistic outflows producing gamma-ray bursts is a long standing open question. One of the main arguments in favor of magnetically dominated outflows is the absence of photospheric component in their broadband time resolved spectra, with such notable example as GRB 080916C.
%In this paper, we perform accurate analysis of time resolved spectra of this GRB and confirm the previous detection of an additional spectral component in GRB 080916C.
Here, we perform a time-resolved analysis of this burst and confirm the previous detection of an additional spectral component.
We show that this subdominant component is consistent with the photosphere of ultrarelativistic baryonic outflow, deep in the coasting regime.  We argue that, contrary to previous statements, the magnetic dominance of the outflow is not required for the interpretation of this GRB. Moreover, simultaneous detection of high energy emission in its prompt phase requires departure from a one-zone emission model.
\end{abstract}

%\maketitle

%\section{TO BE REMOVED}

%\addedD{The following are my notes :}

%ZP09 analysis based on :
%\begin{enumerate}
%    \item featureless spectrum 
%    \item photospheric argument (no BB detected) : use $R_0 = c \delta t$ with $\delta t = 2s$. Which implies acceleration phase for Gamma = 1000 required by compactness argument
%    \item compactness argument (10 GeV photon) : requires emission of GeV photons to be very large. Does not constrain low energy photons : uses the "featureless argument" $\rightarrow$ only one emission region.
%\end{enumerate}
%Peak energy is very high : 8MeV. Can it be explained with photosphere of poynting flux outflow ?
%\begin{enumerate}
%    \item Peak energy 8 MeV ok
%    \item But efficiency only a few percent
%    \item $\rightarrow$ Use of the BB : does not work because too high peak energy
%    \item $\rightarrow$ Featureless spectrum assumption : efficiency at photosphere is only a few percent, which would mean $L \sim 10^{56}$ erg.s$^{-1}$ 
%\end{enumerate}
%\addedD{Damien : end of Damien's notes}

\section{Introduction}

    Gamma-Ray Bursts (GRBs) are exciting astrophysical phenomena observed daily, notwithstanding their sources located at cosmological distances. This is due to their huge power and extremely efficient conversion of energy into radiation. What is certain is that the origin of this phenomenon is the gravitational energy of compact astrophysical sources: possibly of the collapsing core of a massive star or of a neutron star binary. Another certain fact is the electromagnetic radiation detected on Earth. What is not established so far, despite decades of intense research, is the mechanism of energy conversion.
    
    The fireball model, for reviews see \textit{e.g} \citet{1999PhR...314..575P,Pee15,2018pgrb.book.....Z}, assumes that the gravitational energy is converted into heat of electron-positron-photon plasma, loaded with a small fraction of baryons \citep{1990ApJ...365L..55S,1990ApJ...363..218P}. This plasma, being optically thick, expands and converts its thermal energy into kinetic energy of baryons. Various dissipation mechanisms such as internal \citep{1994ApJ...430L..93R} or external \citep{1993ApJ...405..278M} shocks further convert this energy into radiation, which we eventually observe. The fireball model was successful in explaining the main observational features of GRBs. It also predicted two key signatures: a long lasting multiwavelength afterglow following the prompt $\gamma$-ray emission \citep{1997ApJ...476..232M}, and a photospheric emission \citep{1986ApJ...308L..47G,1986ApJ...308L..43P,2000ApJ...530..292M}, released during the plasma expansion when it becomes optically thin, see \citet{2014IJMPD..2330003V,2017SSRv..207...87B,2017IJMPD..2630018P} for reviews on photospheric emission. Photospheric emission was successfully identified in several bursts observed by BATSE \citep{Ryd04,Ryd05,RP09} and more recently by {\it Fermi}-GBM, see e.g. \citet{RAZ10,IRA13,IRA15,LRB15,RLA17,Meng2018,Meng2019,DPR20}.
    
    Despite those successes, several difficulties come with the baryonic fireball model, see e.g. \citet{2011ApJ...726...90Z}. One of them is the low energy conversion efficiency of shocks, being around ten percents\footnote{The efficiency of energy conversion of the gravitational energy into heat is uncertain, but heat to kinetic energy conversion is efficient.} for optimistic parameters \citep{KPS97,Bel00}. Another one is the lack of observed bright thermal emission \citep{DM02,HDM13}. As a possible solution, alternative models have been advanced, which assume that gravitational energy is converted first into magnetic energy and propagates in the form of a Poynting flux \citep{1994MNRAS.267.1035U,1994MNRAS.270..480T,2003astro.ph.12347L}. This magnetic energy is then dissipated, e.g. by magnetic reconnection \citep{SDD01,DS02, 2002A&A...387..714D, 2016MNRAS.459.3635B, BPL17, GU19} or by current driven instabilities \citep{2006A&A...450..887G}, and is eventually converted into radiation. Much less is certain about these magnetic models as first principle calculations appear to be difficult, and their predictions are heavily based on numerical simulations \citep{2017MNRAS.468.3202B,2020MNRAS.499.1356G}. Moreover, magnetically dominated outflows are shown to be characterized by relatively bright photospheric emission with excessively large temperature above 8 MeV \citep{2015ApJ...802..134B}, although this results holds true for a specific model of magnetic energy dissipation, namely magnetic reconnection.
    
    Observationally, the very bright GRB 080916C detected by the Fermi Gamma-ray Space Telescope \citep{2009Sci...323.1688A} has been considered as a "Rosetta stone" for magnetic models. Indeed, \citet{ZP09} argued that the absence (at the time) of detection of a photospheric component in its time resolved spectra, rules out the baryonic model and they suggested a magnetically dominated outflow as an alternative. Even though a thermal component was later discovered in this GRB by \citet{2015ApJ...807..148G}, the conclusion presented by \citet{ZP09} was not reevaluated and many following works take it at faced value. For instance, \citet{2015ApJ...801..103G} proposed that GRB sources produce a hybrid composition of relativistic outflows with arbitrary entropy $\eta$ and magnetization $\sigma$. This model was further applied to GRBs detected by {\it Fermi}-GBM and the outflow parameters, such as the Lorentz factor $\Gamma$ were determined, see \textit{e.g.} \cite{Li2020}. Moreover, \citet{2018NatAs...2...69Z} and \citet{Li2019a} used this model to argue for a change of outflow composition for each different emission episode of GRB 160625B, being initially baryonic and latter magnetic. In fact, the absence of a bright photosphere is often considered as an argument in favor of magnetically dominated outflow \citep{2011ApJ...726...90Z}, or at least that a substantial fraction of the available energy is magnetic \citep{DM02,HDM13,BBB19}.

    %\authorcommentD{This paragraph should be rephrase. The high energy component is loosely defined and it is an issue.}
    In this paper, we perform a detailed spectral analysis of GRB 080916C, and confirm the previous results of \citet{2015ApJ...807..148G} on the presence of an additional component in the time resolved spectra. By interpreting this spectral component as photospheric emission and by applying the method of \citet{2007ApJ...664L...1P} we estimate the Lorentz factor of the baryonic outflow, the nozzle radius $r_0$ and the photospheric radius $r_{\rm ph}$ of the outflow. Our results indicate that observations of GRB 080916C can be naturally accounted for in the standard baryonic fireball model, provided that the high energy emission (photons with energy a few hundreds MeV), detected in this GRB originates not at the photosphere, but at larger radii. This is consistent with the earlier proposals of high energy emission emerging in the external shock, see \textit{e.g.} \citet{2009MNRAS.400L..75K}. Here, we argue that the main assumption behind the conclusions of \citet{ZP09},  is the one-zone approximation for the broadband emission. This assumption, originally supported by the ''featureless'' spectrum, is in fact not required by the data. Once it is released, the observations might be explained in both baryonic and magnetic models.
    
    The paper is organised as follows. In Section \ref{sec:2}, we revisit the argument for magnetic dominance of the outflow. In Section \ref{sec:3}, we perform a time-resolved spectral analysis of GRB 080916C, hereby confirming the results of \citet{2015ApJ...807..148G} on the presence of an additional thermal component. In Section \ref{sec:4}, we determine the outflow parameters within the baryonic fireball model. The discussion and conclusion follow.
    
\section{GRB 080916C and the argument for magnetic dominance}
    \label{sec:2}
    
    \begin{figure*}
    \begin{center}
    \includegraphics[angle=0,scale=0.55]{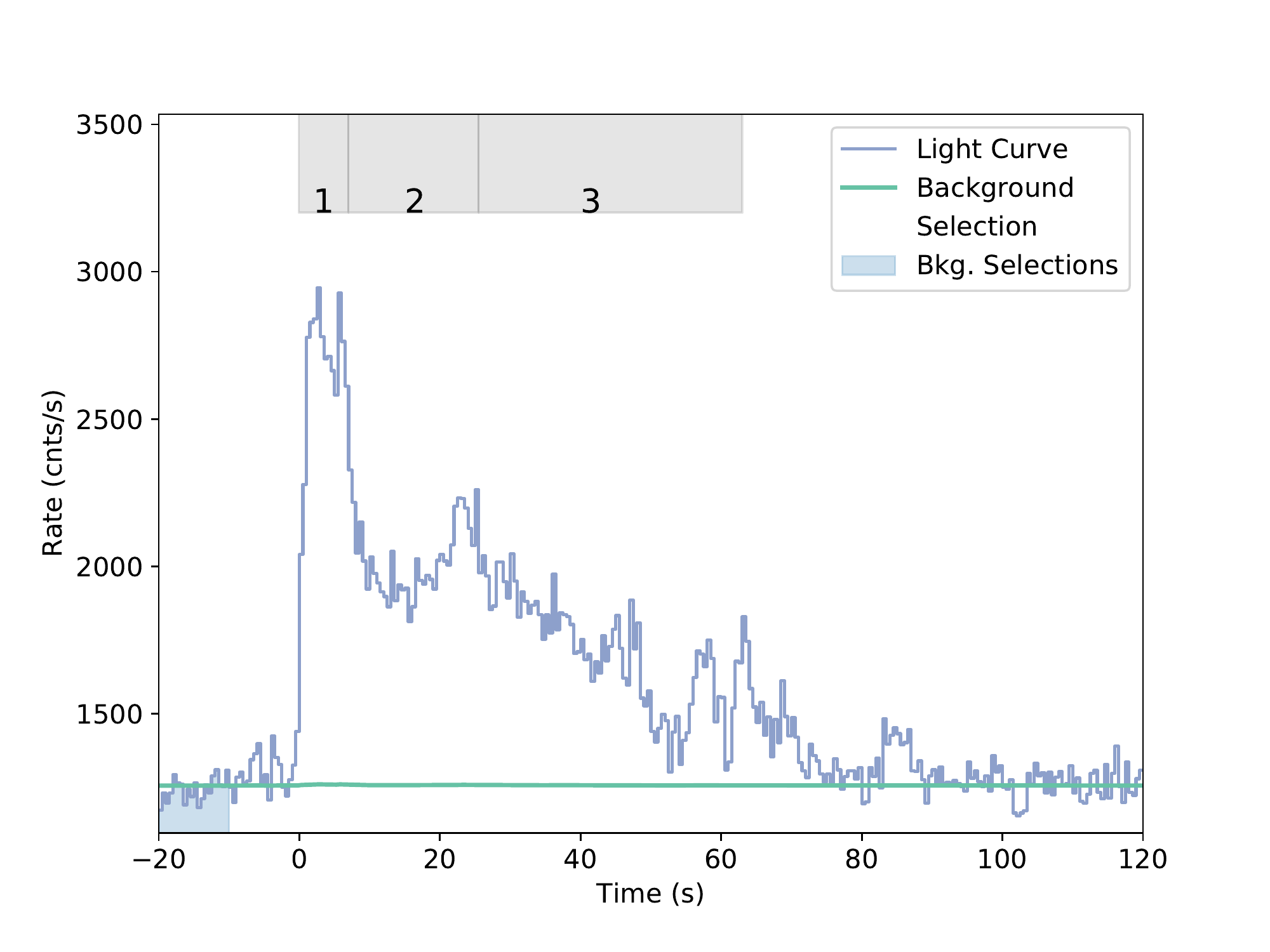}
    \end{center}
    \caption{The prompt emission light curve of GRB 080916C from {\it Fermi}-GBM observed by the NaI4 detector. The regions labeled 1, 2, and 3 corresponds to the time bins in which the spectral analysis is performed.}
    \label{fig:lc}
    \end{figure*}

    On September 16, 2008 at 00:12:45.61 Universal Time (UT),  GRB 080916C was detected by the Gamma-ray Burst Monitor (GBM: 8 keV-40 MeV, \citet{Meegan2009}) and the Large Area Telescope (LAT, 20 MeV-300 GeV, \citet{Atwood2009}) on board the NASA {\it Fermi} Gamma-Ray Observatory. In addition to {\it Fermi} \citep{2008GCN..8251....1H}, the burst was bright enough to also trigger AGILE (MCAL, SuperAGILE, and ACS), RHESSI, INTEGRAL (SPI-ACS), Konus-Wind, and MESSENGER. Long term follow-up observations of the burst were carried out by {\it Swift} \citep{2008GCN..8253....1K} and several ground-based optical telescopes \citep{2008GCN..8257....1C, 2008GCN..8272....1C}. The prompt emission lightcurve exhibits an initial short peak lasting about 15s followed by a longer and smoother episode, see Figure \ref{fig:lc}. The total burst duration is $T_{90} = 63 \pm 0.8$s, corresponding to the time during which 90\% of the burst fluence is accumulated. This burst is very bright, and the fluence between 10-1000 keV reported by GBM term is (6.0272$\pm$0.007)$\times$10$^{-5}$ erg cm$^{-2}$. With a redshift of z = 4.35 $\pm$ 0.15 \citep{Greiner2009}, its isotropic gamma-ray energy is estimated to be $E_{\rm iso } \sim 9\times 10^{54}$ erg, making it one of the most energetic GRB ever observed.

%The detection of \replacedD{high energy}{GeV} photons \replacedD{in}{during} the prompt \replacedD{emission}{phase} of GRB 080916C allowed \citep{2009Sci...323.1688A} to set up lower limits on the Lorentz factor of the outflow, which are $608\pm 15$ and $887\pm 21$ for time intervals $3.58-7.68$ s and $15.87-54.78$ s, respectively. Considering the \addedD{conservative} time variability of 2s of this source, the lower limit on emission radius was computed $R>8.9\times 10^{15}(\Gamma/890)^2(\Delta t/2\,{\rm s})[(5.35)/(1+z)]$ cm for this source. More refined analysis based on compactness argument and an assumption that above the peak the spectrum remains a power law at high energies \citep{ZP09} confirmed these estimates.

%\authorcommentD{Actually \citep{ZP09} used the same approached using the spectral shape.}

    Within the fireball model, the detection of GeV photons during the prompt phase of GRB 080916C implies a large value for the outflow bulk Lorentz factor $\Gamma$. \citet{2009Sci...323.1688A} estimated a lower limit on the Lorentz factor based on the spectral parameters from the Band model (see below) and a conservative variability time of $\Delta t = 2$s. For the time interval $3.58-7.68$ s, a lower limit of $\Gamma_{\rm min} = 887\pm 21$ was found. A similar constrain of $\Gamma_{\rm min} = 608\pm 15 $ was obtained for the time interval $15.87-54.78$s, although the variability time was assumed to be $\Delta t = 20$s. Those values where derived under the  assumption that the radius is linked to the Lorentz factor and the variability time as $R = c \Delta t \Gamma^2$, leading to emission radius of $R \sim 10^{16}$cm. These estimates were then confirmed by \citet{ZP09}.

%\deletedD{Based on these large emission radii and large Lorentz factor limits, as well as variability timescale \citet{ZP09} computed the expected signal from the photospheric emission in baryonic outflow and claimed that this signal should have been well detected. Based on the absence of detection, their argued in favor of a different acceleration mechanism with respect to thermal acceleration $\Gamma\propto r$ inherent in the baryonic model, specifically in favor of magnetic dominance of the outflow, known for slower acceleration $\Gamma\geq r^{1/3}$ \citep{2011ApJ...733L..40M}.}

    %Relying on such a large emission radii and large Lorentz factor estimates, \citet{ZP09} computed the expected flux and temperature from the photospheric emission in baryonic outflow and claimed that this signal should have been well detected. Since for this specific burst, no detection of a thermal component was reported at that time, \citet{ZP09} argued in favor of a different acceleration mechanism with respect to thermal acceleration $\Gamma\propto r$ inherent in the baryonic fireball model, specifically in favor of magnetic dominance of the outflow, known for slower acceleration $\Gamma\geq r^{1/3}$ \citep{DS02,GKS11,2011ApJ...733L..40M, BPL17, GU19}. \authorcommentD{In their paper ZP09 do not mention anything on the acceleration mechanism. In fact they do not change the acceleration mechanism. This needs to be rewriten to the magnetic dominance, and then possibly different acceleration mechanism}

%\authorcommentD{New paragraph version :}
    Relying on such a large emission radii and large Lorentz factor estimates, \citet{ZP09} computed the expected flux and temperature from the photospheric emission in baryonic outflow and claimed that this signal should have been well detected. The key assumption in the computation of \citet{ZP09} is the association between the nozzle radius $r_0$ and the  variability time $\Delta t$, such that $ r_0 = c \Delta t \sim 6 \times 10^{10}$cm, where $c$ is the speed of light. The nozzle radius is the radius from which the outflow starts accelerating freely because of its own internal energy. This assumption leads to a very efficient photospheric emission since the photospheric radius would nearly be equal to the coasting radius $\eta r_0$ for a Lorentz factor $\Gamma \sim 10^3$, as suggested by the observed GeV photons. Since for this specific burst, no detection of a thermal component was reported at that time, \citet{ZP09} argued that the energy be initially dominated by magnetic energy. Although not discussed by \citet{ZP09}, this composition of the flow leads to a different acceleration mechanism with respect to thermal acceleration $\Gamma\propto r$ inherent in the baryonic fireball model, specifically to a slower acceleration $\Gamma\geq r^{1/3}$ typical of magnetic outflows \citep{DS02,GKS11,2011ApJ...733L..40M, BPL17, GU19}, as well as different spectral emission properties.

    Subsequently, a photospheric component was reported in the detailed spectral analysis performed by \citet{2015ApJ...807..148G}. Nevertheless, \citet{2015ApJ...801..103G} argued that the conclusions of \citet{ZP09} remain valid based on the fact that the reported thermal component has a fluence comparable to the limit they used. %This conclusion would be correct for their assumed values of dimensionless entropy $\eta = 470$ and magnetization $\sigma = 20$. Yet, we will demonstrate below that a solution solely based on a baryonic flow can be obtained and therefore baryonic fireball is not ruled out for this GRB.
    However, it should be emphasized that magnetic outflow cannot explain such a subdominant photospheric component either. Indeed, three-body photon non-conserving processes are suppressed, which leads to photon starvation and as a consequence to the shift of temperature to the MeV region \citep{2015ApJ...802..134B}. Besides, the photospheric signal is predicted to be sufficiently bright to be detected. These results are weakly dependent on parameters of the outflow. Hence our main conclusion is that the observed broadband spectrum of GRB 080916C (the dominant non-thermal component and the subdominant black-body) cannot be explained with any model based on a one-zone approximation, i.e. assuming that all multiwavelength emission is produced at the same radial position.

    In what follows, we reanalyse the spectra of GRB 080916C and based on these results, we show that a baryonic model can account for the photospheric component discovered in GRB 080916C, without invoking the hypothesis of magnetic dominance.

\section{Time resolved spectral analysis of GRB 080916C}
\label{sec:3}

\begin{figure*}
\centering
\begin{tabular}{cc}
\includegraphics[width=0.43\textwidth]{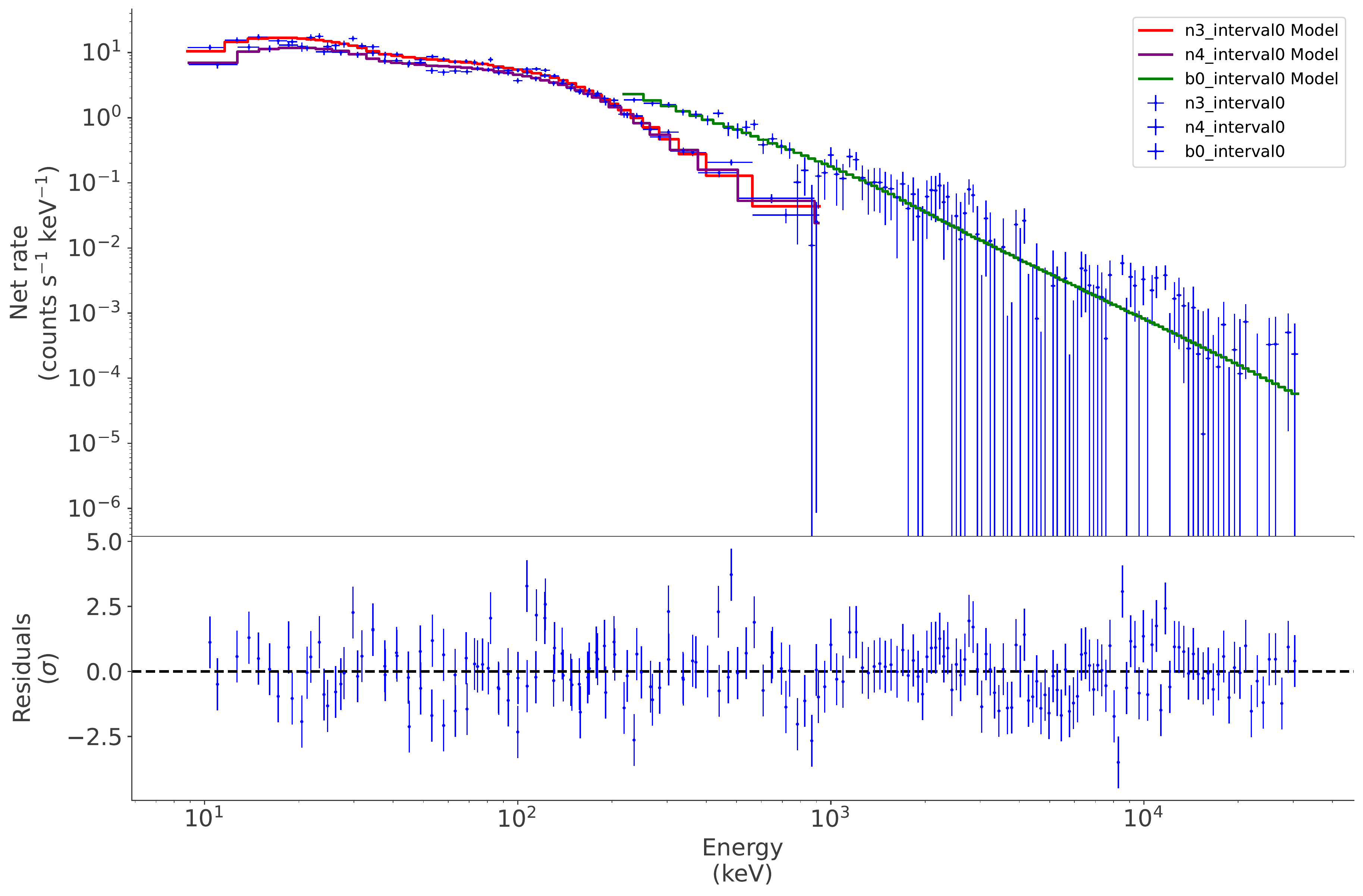} &
\includegraphics[width=0.43\textwidth]{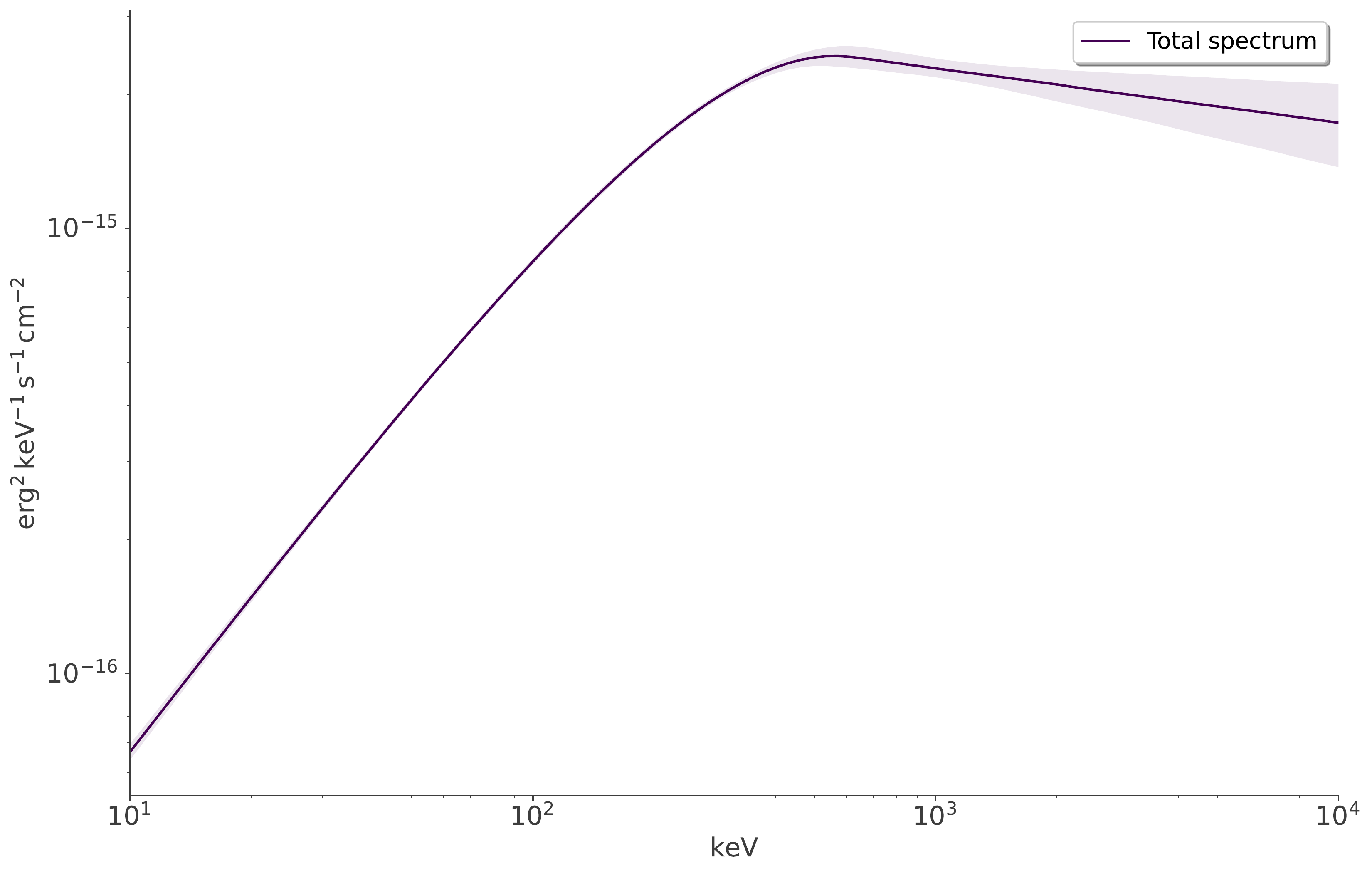} \\
\includegraphics[width=0.43\textwidth]{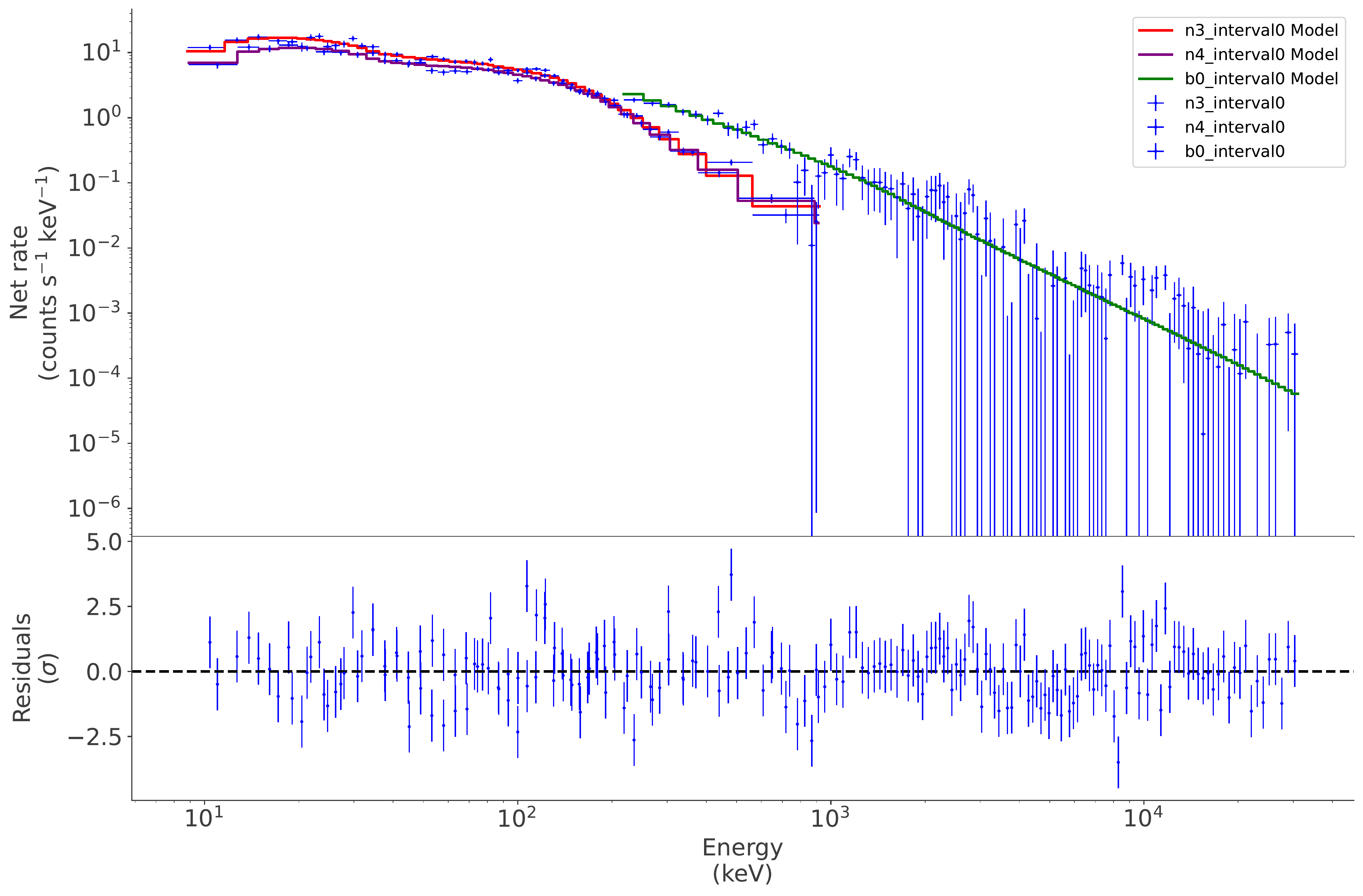} &
\includegraphics[width=0.43\textwidth]{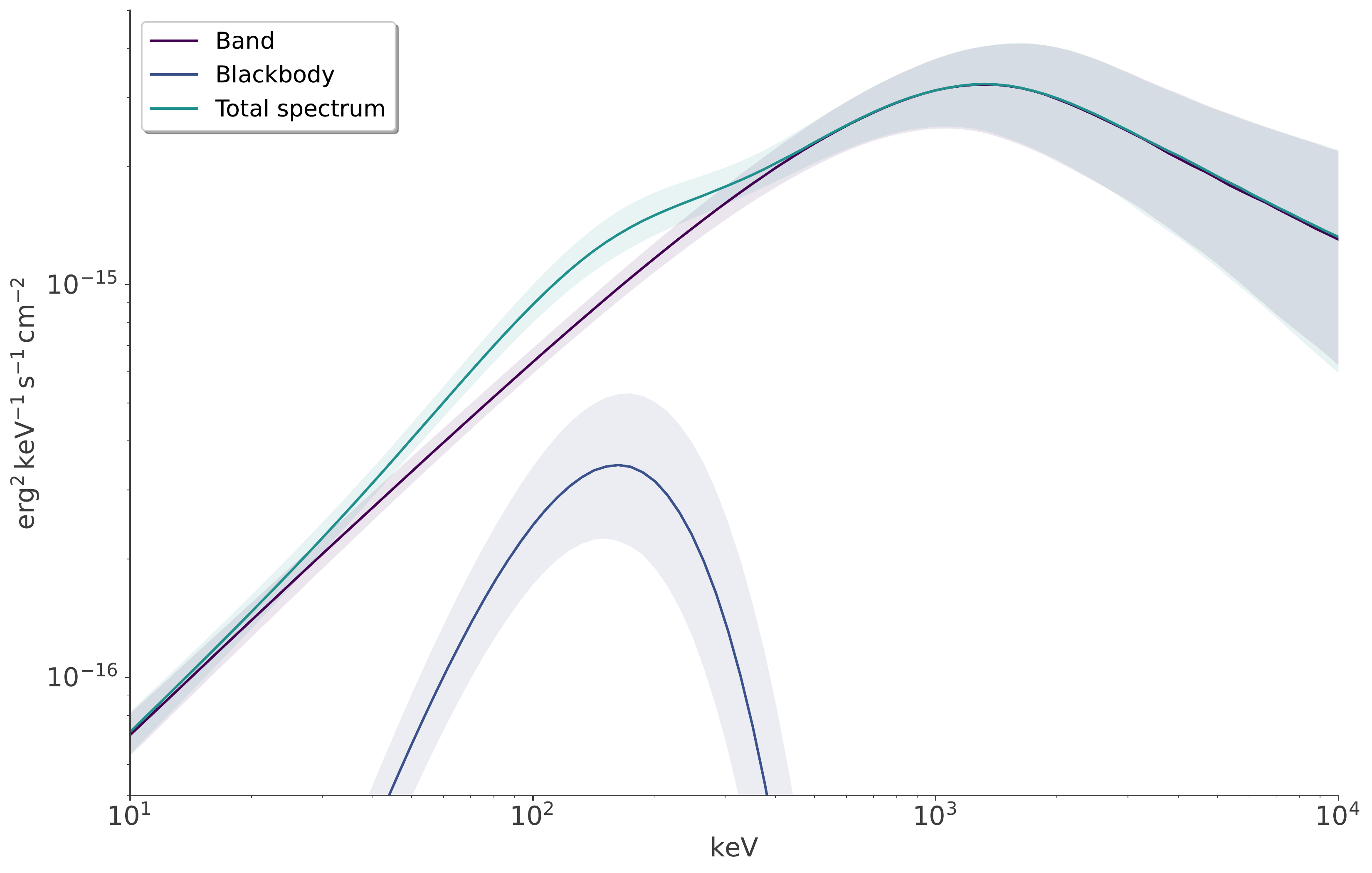} 
\end{tabular}
\caption{The prompt emission spectrum between 0 and 6.98 s of GRB 080916C. Left : count spectrum. Right : $\nu F_\nu$ and model uncertainty. Top : Band model only. Bottom : Band plus black-body.}
\label{fig:spectralfit}
\end{figure*}

%%   OLD TABLE ONE WITH ALL SPECTRAL MODEL : REMOVED
%\begin{table}
%\renewcommand\arraystretch{1.2}
%\centering
%\begin{tabular}{cclccc}
%\hline
%Instrument & Time & Model & Log(Likelihood)& AIC & BIC \\
%&(s)&&&& \\
%\hline
%\emph{Fermi}-GBM&-1-62.977($T_{90}$)&PL&3381&6766&6774\\
%&&BB&3877&7758&7765\\
%&&CPL&2914&5835&5846\\
%&&Band&2904&5816&5831\\
%&&PL+BB&2943&5894&5909\\
%&&PL+CPL&2905&5820&5839\\
%&&CPL+BB&2902&5814&5833\\
%&&\underline{Band+BB}&2898&5809&5831\\
%&&\underline{CPL+BB+PL}&2898&5810&5836\\
%\hline
%\end{tabular}
%\caption{\authorcommentD{Can we just remove this table ? It is not giving any information}\deletedD{The statistics of the time-integrated prompt spectrum of GRB080916C. Data is from {\it Fermi}-GBM and the duration of $-1-62.977$~s ($T_{90}$) and three GBM detectors (nai3, nai4 and bgo0) are used. Statistically, the Band+BB or CPL+BB+PL model gives the best fit.}}
%\label{tab:promptModel}
%\end{table}

%\authorcommentD{Why so many models ? Can we stick to 1- CPL 2- Band, 3- Composite with BB ? Also why do we even need time integrated spectrum ???}

Our procedure for performing the spectral analysis for GRB 080916C involves the following steps:
(1) Using 3ML (the Multi-Mission Maximum Likelihood Framework, see \citealt{Vianello2015}), we follow the standard practices provided by the {\it Fermi} team to carry out the spectral analysis \citep{Burgess2019,Li2019a,Yu2019,DPR20,Li2021b}, see also  \citet{GBP12,GGW14,YPG16,Li2021a}, that is to say we select a background region (-20 to -10 s, 120 to 140 s) and a source interval (0 to 63 s) in the sodium iodide detectors NaI 3 and NaI 4, and in the bismuth gallium oxide BGO 0. The background is fitted independently in each detector with a polynomial, which order is determined via a likelihood ratio test, (2) Since it was shown that time-resolved spectral analysis is critical for identifying and physically interpreting extra components, see e.g. \citet{BR15}, we bin the time-tagged events light-curve with the Bayesian blocks algorithm \citep{SNJ13}, and we compute the significance (S; \citealt{Vianello2018a}) for each time bins. For some of those time bins, the significance is low and so is the signal over noise ratio. This means that not all parameters of the fits converge. Therefore, we re-bin the data by collocating time bins until each model parameter can converge. This procedure leads to three time bins. (3) We finally perform a time-resolved spectral analysis using two models : (i) the Band model supporting the original analysis \citep{Meegan2009,Atwood2009,ZP09}, (ii) the Band model plus black-body. \citet{Guiriec2015a} used a three component model composed of a  cutoff power-law, a power law and a black-body. We tried to use this model as well, but could not constrain the high energy power-law, most likely because our analysis does not include LAT data. Yet, the black-body component we find is consistent with the one found in \citet{Guiriec2015a}.

To perform the data analysis, we used the following priors for the Band function :
\begin{itemize}
    \item Band normalisation $\pi(K)$ = LogUnif $\left ( 10^{-10} \right.$, $\left.10^3 \right )$ in cm$^{-2}$ keV$^{-1}$s$^{-1}$,
    \item Band low energy index $\pi(\alpha)$ = Unif $\left ( -1.5 \right.$, $\left.3 \right )$,
    \item Band high energy index $\pi(\beta)$ = Unif $\left ( 5 \right.$, $\left. -1.6 \right )$,
    \item Band break energy $\pi(x_c)$ = LogUnif $\left ( 10 \right.$, $\left. 10^4 \right )$ in keV,
\end{itemize}
while the priors for the additional black-body were set to 
\begin{itemize}
    \item temperature $\pi(k_B T)$ = LogUnif $\left ( 1 \right.$, $\left.10^2 \right )$,
    \item normalisation $\pi(K_{\rm BB})$ = LogUnif $\left ( 10^{-10} \right.$, $\left.10^3 \right )$ in cm$^{-2}$ keV$^{-1}$s$^{-1}$.
\end{itemize}
We have tried different priors and find essentially the same results.

We found that a black-body component is detected only in the first time bin. This is similar to the results of \citet{Guiriec2015a}. In their analysis they found a black-body which fluxes decreases with time. Table \ref{tab:band_only} gives the value of all parameters obtained by fitting the Band  model to GBM data and Table \ref{table:080916C_BBB} gives the value of the parameters for the Band plus black-body model in the first time bin. Since a black-body is not detected in the second and third time bin, we do not report the fit results of the second and third time bin for the Band plus black-body model. Figure \ref{fig:spectralfit} shows the count spectra and the $\nu F_\nu$ spectrum with model uncertainty. The spectrum is composed of a dominant non-thermal component, with peak energy at $\sim 1.5$MeV and an additional black-body with temperature $ k_B T\sim 40 $keV, where $k_B$ is the Boltzmann constant.

\section{Determination of the outflow's parameters}

\label{sec:4}

Assuming that the additional spectral component responsible for the bump around $100$ keV is of photospheric origin in a baryonic outflow, it is possible to estimate the Lorentz factor and the nozzle radius $r_0$ of the outflow directly from observations \citep{2007ApJ...664L...1P}. In the
ultrarelativistic regime one defines%
\begin{equation}
\mathcal{R}\equiv\left(  \frac{F_{obs}^{BB}}{\sigma_{SB}T_{obs}^{4}}\right)
^{1/2}=\zeta\frac{\left(  1+z\right)  ^{2}}{d_{L}}\frac{R}{\Gamma},
\label{Rcal}%
\end{equation}
where $R$\ is the emission radius, $d_{L}$\ is the luminosity distance, $\zeta$\ is a numerical factor of order unity. Following \citet{2007ApJ...664L...1P}, $\zeta=1.06$ is chosen for the estimates below. In the derivation, it is assumed that the outflow is coasting at ultrarelativistic speed\footnote{The reader interested in the derivation in other regimes relevant for GRBs is refered to \citet{BI14,VS20} and \citet{ZWL21}.}, $\Gamma=\eta\gg1$ and that the total luminosity is given by $L=4\pi d_{L}^{2}YF_{obs}$\, where $Y$ is the fraction of total energy to the energy emitted in X and $\gamma$-rays. For GRB 080916C, this is a fair approximation since the thermal component only represents a few percent of the total emitted energy. The nozzle radius is%
\begin{equation}
r_{0}=\frac{4^{3/2}d_{L}}{\xi^{6}\zeta^{4}\left(  1+z\right)  ^{2}}%
\mathcal{R}\left(  \frac{F_{obs}^{BB}}{YF_{obs}}\right)  ^{3/2}, \label{R0det}%
\end{equation}
where $\xi=1.48$ for the steady wind approximation \citep{2007ApJ...664L...1P}. Since the outflow is in the coasting regime to a good approximation, the Lorentz factor is
\begin{equation}
\Gamma=\left[  \zeta\left(  1+z\right)  ^{2}d_{L}\frac{\sigma YF_{obs}%
}{m_{p}c^{3}\mathcal{R}}\right]  ^{1/4}. \label{etaPeer}%
\end{equation}

Following the formalism from \citet{2007ApJ...664L...1P}, we determine the physical characteristics of the outflow producing the photospheric emission. These results are summarized in Table \ref{table:080916Cres}. Remarkably, the value of the Lorentz factor $\Gamma \sim 1000$ is consistent with the estimates obtained from the compactness argument, based on observation of $\sim 10$GeV photons. Besides, given the relatively small value of the nozzle radius $r_0\sim 10^6$ cm and the large value of the photospheric radius $r_{ph}\sim 10^{12}$ cm, the outflow appears to be deep in the coasting regime. This explains the relative weakness of the photospheric signal, which accounts for only $\sim$ 5 per cent of the total flux. This is in contradiction with the assumption of \citet{ZP09}, who assumed that $r_0$ be given by the variability time, see Section \ref{sec:2}. This explain why the photospheric signal was over predicted in their analysis.

%considered the photospheric radius coincident with the saturation radius, which we find instead much smaller $r_s\sim 10^9$ cm. \replacedD{Hence the reason the photospheric signal was overpredicted there.}{This explain why the photospheric signal was over predicted in their analysis.}

    \section{Discussion and conclusions}

Having identified the subdominant spectral component in the spectrum of GRB 080916C with the photosphere at radius $\sim10^{12}$ cm, one has to deal with the lower limit of emission radius $R>10^{16}$ cm of high energy photons in the GeV energy range. Clearly, such emission cannot originate at the photosphere. Therefore we identify the dominant spectral component \citep{2009Sci...323.1688A,2015ApJ...807..148G} with nonthermal source, probably due to external shock \citep{2009MNRAS.400L..75K} or other dissipation mechanism. This is similar to the assumption of \citet{BBR16} in the analysis of GRB 141028A interpreted in the framework of an external shock. Similar arguments to the one discussed above are applicable to other GRBs, in particular to GRB 160625B, which were interpreted as evidence of transition from baryonic fireball to a magnetic dominance \citep{2018NatAs...2...69Z}.

These considerations show that a one zone radiation model fails for a baryonic fireball model. But clearly the same argument applies to the magnetic outflow model as well. However, the subdominant photospheric component with temperature $40$ keV cannot be produced in magnetized outflows, as one expects much higher temperatures $T>8$ MeV \citep{2015ApJ...802..134B}. Therefore, contrary to the statement in \citet{ZP09}, we interpret the observation of the photospheric component as evidence against the magnetic dominance of the outflow in GRB 080916C. In fact, if the dominant component is interpreted as a synchrotron emission in external shock, it appears to be consistent both with expected flux decay and with magnetic field amplification in the circumstellar medium \citep{2009MNRAS.400L..75K}.

 %Indeed, the photospheric signal with $30$ keV detected in its first emission episode disappears in subsequent episodes. Instead, high energy photons are simultaneously detected in the second and third episodes. What is not typical in this case is long quiescent periods between the episodes, which might be attributed to episodes of activity of the central source. %  \authorcommentD{I am not sure what you are trying to say here.}

Our study underlines once more the intrinsic difficulties of using phenomenological functions in spectral analysis to give firm conclusion on the emission mechanism of GRBs or the outflow content (baryonic vs magnetic). Both models used for the spectral analysis performed in this paper provide an adequate description of the data but lead to fully incompatible conclusions on the nature of the outflow of GRB 080916C. Similar issues were encountered while comparing fitted results from the Band function to expectations from the synchrotron mechanism \citep{AB15,YVG15,Bur17}. This issue is partially solved by using physical models \citep{ALN15,VGG17,ALA19,BBB19} and could be further lifted by using additional information such as polarization, see \textit{e.g.} \citet{BKB19,RGK20} or spectral evolution \citep{BBR16}.

In conclusion, we have performed spectral analysis of GRB 080916C revealing the presence of an additional subdominant spectral component, which we, following \citet{Guiriec2015a} interpret as due to emission from the photosphere of a relativistic outflow. The presence of this photospheric emission with temperature around $40$ keV implies that the outflow has bulk Lorentz factor about $10^3$ when it becomes transparent to radiation. This is in agreement with the Lorentz factor estimate based on the compactness argument. The observation of high energy emission in this burst implies a large emission radius, which is consistent with an external shock origin \citep{2009MNRAS.400L..75K}. Altogether these data imply that a single-zone radiation model cannot be used for interpreting the spectra of this GRB. Although it could still be a viable alternative scenario, provided its spectrum can match the observations, we find that magnetic dominance of the outflow is not required by the data.

\begin{deluxetable}{cccccccccc}
\centering
%\rotate
\tabletypesize{\scriptsize}
\tablecaption{Time-resolved spectral parameters for the Band model.}
\tablehead{
%\specialrule{0em}{3pt}{3pt}
\colhead{$t_{\rm start}$$\sim$$t_{\rm stop}$}
&\colhead{$S$}
&\colhead{$\alpha$}
&\colhead{$\beta$}
&\colhead{$x_{\rm c}/10^2$}
&\colhead{$K/10^{-2}$}
&\colhead{Flux /$10^{-6}$}
&\colhead{likelihood}
&\colhead{AIC}
&\colhead{BIC}
}
\colnumbers
\startdata
\hline
0.00$\sim$6.98 & 80.93 & -0.82 $\pm 0.15$  & -2.14$^{+0.1}_{-0.09}$ &5.6 $\pm 0.5$ & 4.07$\pm 0.15$ & 6.07$\pm 0.3$ &2163 &  4333 & 4350 \\
6.98$\sim$25.50 & 66.64 & -0.97$\pm 0.03$ & -1.99$^{+0.09}_{-0.08}$ & 4.4$\pm 0.5$ & 2.2 $\pm 0.1$ &  2.79 $\pm 0.2 $ & 2626 & 5260 & 5275 \\
25.50$\sim$63.00 & 48.80 & -0.97$\pm 0.4$ & -3.3$^{+0.9}_{-1}$ & 4$^{+0.4}_{-0.5}$ & 1.33 $\pm 0.07$ & 0.93$^{+0.19}_{-0.09}$  & 2956 & 5920 & 5935 
\enddata
\tablecomments{The fluxes unit is erg/cm$^2$/s and the normalisation $K$ and $K_{BB}$ have units cm$^{-2}$ keV$^{-1}$s$^{-1}$. }
\label{tab:band_only}
\end{deluxetable}

%\clearpage
\begin{deluxetable}{cccc|cc|cc|ccc}
\centering
%\rotate
\tabletypesize{\scriptsize}
\tablecaption{Parameters for the Band plus black-body model in the first time bin (0-6.98s).}
\tablehead{
\specialrule{0em}{3pt}{3pt}
\multicolumn{4}{c}{Band component} &
\multicolumn{2}{c}{BB component} &
\multicolumn{2}{c}{Flux} &
\multicolumn{3}{c}{Statistic}\\
%\cmidrule(lr){3-4}  \cmidrule(lr){5-5} \cmidrule(lr){6-6}
\colhead{$\alpha$}
&\colhead{$\beta$}
&\colhead{$x_{\rm c}/10^3$}
&\colhead{$K/10^{-2}$}
&\colhead{k$T$}
&\colhead{$K_{\rm BB}/10^{-4}$}
&\colhead{$F_{\rm BB}/10^{-7}$}
&\colhead{$F_{\rm Band}/10^{-6}$}
&\colhead{likelihood}
&\colhead{AIC}
&\colhead{BIC}\\
\hline
}
\colnumbers
\startdata
$-1.02^{+0.18}_{-0.2}$  &  $-2.62^{+0.35}_{-0.33}$  & $1.33 \pm 0.3 $ & $2.71 ^{+0.18}_{-0.2}$ & $41^{+4}_{-3.2}$ & $ 4 \pm 4 $ & $ 3^{+1.6}_{-1.1} $ & $ 6.4^{+1.5}_{1.4}$ & 2144 & 4300 & 4323 
\enddata
%\vspace{3mm}
\tablecomments{The fluxes unit is erg/cm$^2$/s and the normalisation $K$ and $K_{BB}$ have units cm$^{-2}$ keV$^{-1}$s$^{-1}$. }
\label{table:080916C_BBB}
\end{deluxetable}

\begin{deluxetable}{cccccc}
\centering
%\rotate
\tabletypesize{\scriptsize}
\tablecaption{Photosphere properties of GRB 080916C \label{table:080916Cres}}
\tablehead{
\colhead{$t_{\rm start}$$\sim$$t_{\rm stop}$}
&\colhead{$\Re$}
&\colhead{$\Gamma$}
&\colhead{$r_{0}$}
&\colhead{$r_{s}$}
&\colhead{$r_{\rm ph}$}\\
\hline
&(10$^{-19}$)&&(10$^{6}$ cm)&(10$^{9}$ cm)&(10$^{12}$ cm)
}
\colnumbers
\startdata
%\hline
%CPL+BB+PL\\
%\hline
%-0.09$\sim$6.98&4.53$\pm$0.27&950$\pm$24&6.06$\pm$1.20&5.76$\pm$1.14&1.77$\pm$0.56\\
%6.98$\sim$25.50&1.86$\pm$0.26&978$\pm$39&1.11$\pm$0.40&1.08$\pm$0.41&0.75$\pm$0.21\\
%25.50$\sim$63.00&3.12$\pm$0.45&718$\pm$30&1.64$\pm$0.62&1.18$\pm$0.46&0.92$\pm$0.45\\
%-1.00$\sim$63.00&2.65$\pm$0.24&853$\pm$24&1.72$\pm$0.40&1.47$\pm$0.35&0.93$\pm$0.26\\
%\hline
%Band+BB\\
\hline
-0.09$\sim$6.98&3.01$\pm$0.25&1080$\pm$28&2.63$\pm$0.60&2.84$\pm$0.66&1.34$\pm$0.27\\
%6.98$\sim$25.50&2.00$\pm$0.22&980$\pm$38&2.28$\pm$0.75&2.23$\pm$0.73&0.81$\pm$0.27\\
%25.50$\sim$63.00&1.39$\pm$0.35&899$\pm$72&0.71$\pm$0.38&0.64$\pm$0.35&0.52$\pm$0.26\\\hline
%-1.00$\sim$63.00&1.88$\pm$0.18&955$\pm$32&1.75$\pm$0.48&1.67$\pm$0.45&0.74$\pm$0.20\\
\enddata
\vspace{3mm}
Note---parameter $\Re$ (Col.2), bulk Lorentz factor $\Gamma$ (Col.3), the nozzle radius of the outflow $r_0$ (Col.4), saturation radius $r_{s}$ (Col.5), and the photospheric radius $r_{\rm ph}$ (Col.6).
\label{tab:photosphere}
\end{deluxetable}

\begin{acknowledgments}

We thank Felix Ryde for useful discussions. This research made use of the High Energy Astrophysics Science Archive Research Center (HEASARC) Online Service at the NASA/Goddard Space Flight Center (GSFC). DB acknowledges support from the European Research Council via the ERC consolidating grant $\sharp$773062 (acronym O.M.J.).

\end{acknowledgments}

\end{document}